\documentclass[sd,amsmath,amssymb,reprint,graphicx]{revtex4-1}

\usepackage{graphicx}
\usepackage{dcolumn}
\usepackage{bm}
\usepackage{siunitx}

\draft 

\begin{document}

\preprint{AIP/123-QED}


\title{Faraday cage angled-etching of nanostructures in bulk dielectrics} 



\author{P. Latawiec}
\author{M. J. Burek}
\author{Y.-I. Sohn}
\author{M. Lon\v{c}ar}
\affiliation{John A. Paulson School of Engineering and Applied Science, Harvard University}


\date{\today}

\begin{abstract}
For many emerging optoelectronic materials, heteroepitaxial growth techniques do not offer the same high material quality afforded by bulk, single-crystal growth. However, the need for optical, electrical, or mechanical isolation at the nanoscale level often necessitates the use of a dissimilar substrate, upon which the active device layer stands. Faraday cage angled-etching (FCAE) obviates the need for these planar, thin-film technologies by enabling in-situ device release and isolation through an angled-etching process. By placing a Faraday cage around the sample during inductively-coupled plasma reactive ion etching (ICP-RIE), the etching plasma develops an equipotential at the cage surface, directing ions normal to its face. In this Article, the effects Faraday cage angle, mesh size, and sample placement have on etch angle, uniformity, and mask selectivity are investigated within a silicon etching platform. Simulation results qualitatively confirm experiments and help to clarify the physical mechanisms at work. These results will help guide FCAE process design across a wide range of material platforms.
\end{abstract}


\maketitle 


\section{Introduction}

High-quality heteroepitaxially grown substrates, from III-V semiconductors such as GaAs/AlGaAs to the readily available silicon-on-insulator (SOI) platform, form but a short section of the list of materials with attractive electro-optical, mechanical, piezoelectric, or other properties useful in nano-devices. These thin film structures, on the other hand, enable the necessary three-dimensional architectural control required for the isolation of the active layer from the bulk substrate. For nanomechanics, access to the mechanical degrees of freedom often requires the physical separation of structures from the host, such as in suspended beams, cantilevers, and membranes\cite{Ekinci2005}. Nanophotonics, likewise, requires light confinement due to either total internal reflection or distributed Bragg reflection\cite{Joannopoulos2008}. Although a heterolayer structure composed of materials with distinct photonic or phononic properties can suffice to make these devices, a common route is to go one step further and remove (sacrifice) the layer immediately underneath the device with a selective, isotropic etch, thereby suspending the structure in air. For the most part, this design constraint severely restricts the use of a wide range of materials, like the metal oxides (LiNbO$_3$, BaTiO$_3$), group IV (SiC or diamond), III-V (GaN), II-VI semiconductors (PbS, CdS), and other high quality single crystals (quartz, sapphire) due to the difficulty of growing thin films ($\sim$1 $\mu$m) on host substrates at requisite quality and crystallinity. Nonetheless, the push for chip-scale devices made out of these materials has not faltered, with recent demonstrations of thin film technologies (i.e. crystal-ion slicing for complex metal oxides\cite{Levy1998, Izuhara2002, Izuhara2003} and diamond\cite{Zalalutdinov2011, Liao2010}, alternative preparation from the bulk\cite{Hausmann2012, Xiong2011}) enabling the fabrication of nanophotonic devices\cite{Wang2007, Wang2014, Latawiec2015}, albeit at lower quality than commercially available bulk substrates. 

Outside of the thin film paradigm, several pattern transfer techniques with three-dimensional control have been developed.  A commonality across a number of these methods (which this work shares) is the modification of the ion angular distribution during etching. This includes techniques such as Reactive Ion Beam Etching (RIBE)\cite{Lee1979, Cheng1996}, ion-sheath sculpting\cite{Takahashi2009a}, passivation gas flow and DC bias control\cite{Walavalkar2013a}, and focused ion beam etching\cite{Babinec2010, Bayn2011, Zhong2015}. A second class of methods relies on an anisotropic etch followed by a second isotropic etch, where the layer of interest is shielded from etching by a protective coating\cite{MacDonald1996}. Although this technique requires favorable etch chemistries, it has recently seen extension to new platforms like diamond\cite{Khanaliloo2015, Khanaliloo2015b}.

\begin{figure}[!htbp]
\includegraphics{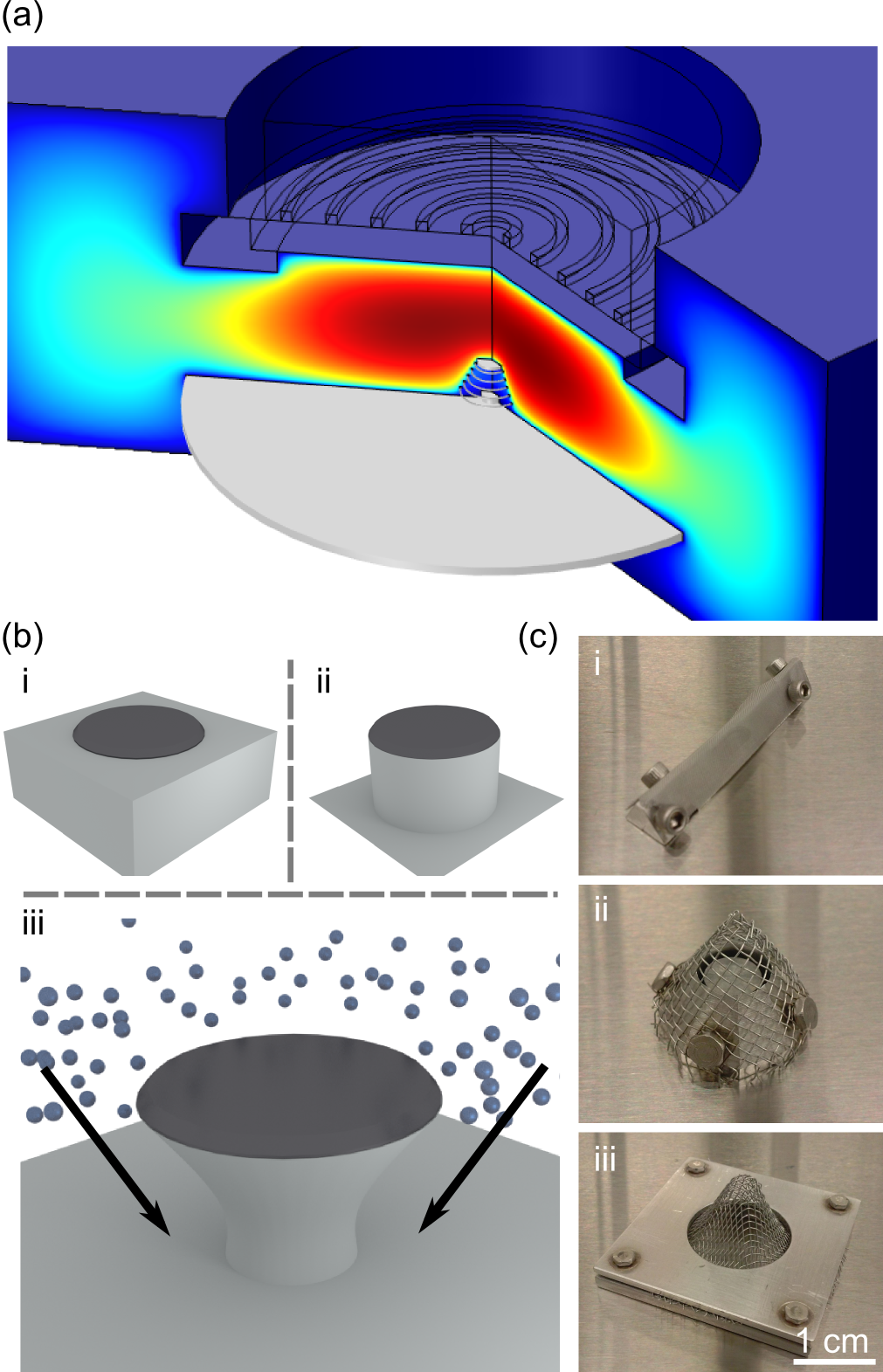}
 \caption{\label{IntroFig} (Color online) Overview of Faraday cage angled-etching (a) 3D simulation of reactor chamber with an argon plasma and a central Faraday cage (details in text). (b) Schematic of angled-etching. The ions, directed by the equipotential on the cage boundaries, are incident upon the sample at an angle. The etch mask defines the shape of the structure. (c) Examples of different cages used for angled-etching, including (i) a triangular cage with a fine mesh (ii) a wrapped cage with a coarse mesh and (iii) a molded cage with a medium mesh, with a shape maintained by retaining plates.}
\end{figure}

In this Article, we study the physics and ion dynamics of a new etching configuration - Faraday cage angled-etching (FCAE). Specifically, a Faraday cage is placed inside an ICP-RIE etch chamber during processing, with the sample placed therein. After the plasma is struck, an equipotential develops at the cage boundaries, resulting in a field-free region inside the cage with electric fields pointing normal to its faces\cite{Boyd1980, Cho1999, Cho, Jeong2009, Burek2012b}. Ions entering from the plasma then get directed toward the sample at an oblique angle of incidence, undercutting the structure and freeing it from the substrate upon completion (Fig. \ref{IntroFig}(b)), effectively fashioning a device from the bulk. Building on initial demonstrations in diamond single crystals\cite{Burek2012b, Bayn2014}, free-standing mechanical cantilevers are shown in silicon and quartz. The effects of varying cage parameters on silicon angled-etching are observed, driving cage optimization. Finally, multiphysics simulations of the dynamics of FCAE elucidate the physical processes affecting etch performance.

\section{Fabrication Methods}

\subsection{Faraday Cage Construction}
Faraday cages built for etching typically consisted of an aluminum base plate with an aluminum mesh arranged around the sample etching region in the shape of a cone or a triangular prism, though in principle other geometries may be interesting\cite{Cho1999, Jeong2009}. Fig. \ref{IntroFig}(c) shows several different cage constructions. Meshes (TWP Inc.) used in the experiment had wires of $\sim$400 $\mu$m diameter at a $\sim$1.6 mm pitch (coarse), $\sim$250 $\mu$m diameter and $\sim$635 $\mu$m pitch (medium), or $\sim$50 $\mu$m diameter and $\sim$125 $\mu$m pitch (fine). Coarse and medium sized meshes could be pressed and molded into shape, while fine meshes required some underlying structure to provide support. The cages were constructed in a variety of sizes, with the only significant limitation being the height of the etching tool's load-lock (the smallest cage was $\sim$7 mm tall and the largest $\sim$18 mm).

Fig. \ref{IntroFig}(c,i) shows a simple, folded mesh over a base support, and Fig. \ref{IntroFig}(c,ii) shows a coarse meshed wrapped around a large aluminum base. Panel (iii) shows a molded-mesh cage design. The mesh is created by pressing it against two custom metal dies, although it can also be done by hand. Because of the modularity of this design it is easy to swap out meshes constructed for different etch angles. The retaining ring around the mesh (Fig. \ref{IntroFig}(c,iii)) provides stability and reduces movement during loading and unloading; likewise, directly affixing the mesh to the carrier wafer via an etch-compatible adhesive maintains angled-etching functionality and placement stability.

Angled-etching of devices follows a process flow similar to standard lithography\cite{Burek2012b}. First, a hard mask is defined either through photolithography or electron beam lithography (Fig. \ref{IntroFig}(b, i)). Once defined, the sample is etched vertically in order to ensure clearance from the substrate (Fig. \ref{IntroFig}(b, ii)). After this step, the sample is placed inside a Faraday cage and etched at an angle, often with the same recipe used to etch vertically (schematically shown in Fig. \ref{IntroFig}(b, iii)). Typically, the extent of etching is recorded via scanning electron microscopy (SEM) in order to time the release of the structure precisely. This becomes especially important for devices where thin supporting regions underneath the device need to be maintained\cite{Burek2014, Latawiec2015b}. After this step is run to completion, the mask is removed. 

\subsection{Diamond FCAE}

Diamond angle-etched devices were first demonstrated for quantum photonics applications\cite{Burek2012b, Hausmann2013a, Bayn2014} and have since shown promise as nanomechanical resonators\cite{Burek2013} and high quality factor optical cavities\cite{Burek2014}. For completeness, we describe the process here. A smooth, polished ($<$1 nm rms roughness) diamond surface is first cleaned in a refluxing mixture of equal parts perchloric, nitric, and sulfuric acid\cite{Atikian2014}. After this, it is placed in a boiling piranha mixture (3:1 sulfuric acid to hydrogen peroxide) before being rinsed and sonicated in solvent, then dried. A thin layer of titanium ($\sim$15 nm) is then deposited on the surface. This helps with resist adhesion and charge compensation during the electron beam lithography. A negative-tone electron beam resist is spun on (FOx-16, hydrogen silsesquioxane, HSQ) to a thickness of $\sim$1 $\mu$m. 

After the resist is exposed and developed, the titanium layer is first removed with a brief Ar/Cl$_2$ plasma etch (Unaxis Shuttleline) at forward power of 250 W and RF power of 400 W at 8 mTorr. This ensures the smooth removal of the titanium underlayer. The sample is then etched with an oxygen chemistry flowing at 50 sccm and maintained at 10 mTorr. The forward and RF powers are held at 100 W and 700 W, respectively. Once the vertical etch has been completed, the sample is placed in the appropriate Faraday cage. The same oxygen etch is run, except for an additional slow flow (2 sccm) of either Ar or Cl$_2$ to mitigate micromasking\cite{Burek2012b}. Once the etch reaches completion, the mask is removed in HF. For sensitive applications, critical point drying can increase the device yield.

\begin{figure}
\includegraphics{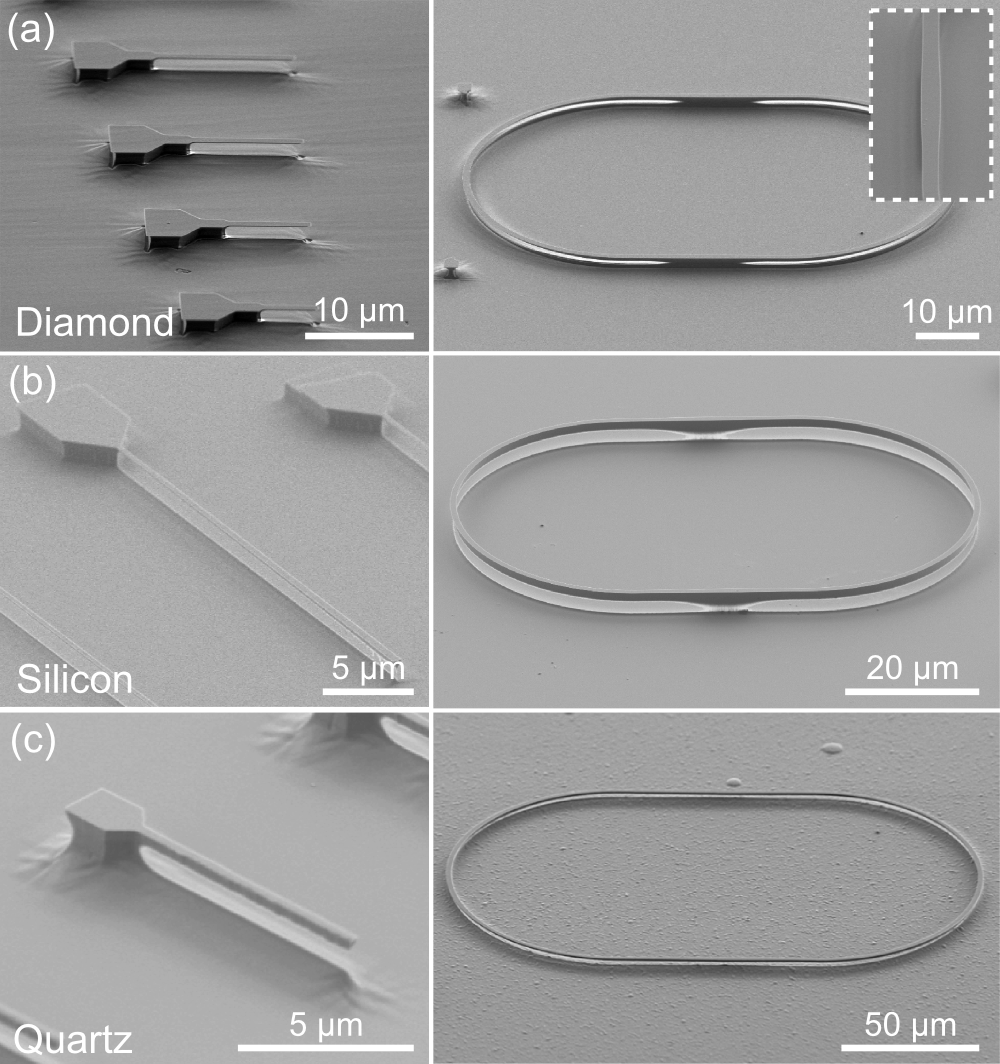}
 \caption{\label{EtchFig} Angled-etched devices (left, nanomechanical cantilevers, right, optical resonators) in different materials. (a) Structures etched in diamond, following the recipe in the text. The lines in the substrate visible on the cantilever device are from the coarse cage used to etch, which impacts the ion trajectory. The optical resonators are supported by flared-out sections in the mask, which transfer to pedestals for an appropriately-timed etch.\cite{Burek2014}. Silicon (b) and quartz (c) cantilevers, $\sim 40 \mu \textrm{m}$ and $\sim 10  \mu \textrm{m}$ long, respectively. All the optical resonators pictured support modes in the telecom wavelength range.}
\end{figure}

\subsection{Silicon FCAE}

Free-standing silicon devices are important in a number of applications and are already made possible by the readily available silicon-on-insulator platform. However, this approach may be inadequate for some select uses, including mid-infrared photonics\cite{Soref2010, Lin2013b, Lin2013c} or mechanical resonators with small vertical cross-section\cite{Llobet2014}. Angled-etching of silicon can circumvent the material thickness restrictions inherent with SOI technology, realizing structures with both lateral and vertical dimensions on the order of tens of nanometers\cite{Latawiec2015b}. Angled-etching of silicon cantilevers (Fig. \ref{EtchFig}(b)) was based on an SF$_6$/C$_4$F$_8$ plasma chemistry (STS MPX/LPX RIE). Two mask materials were shown to have sufficient selectivity: a lifted-off sputtered alumina mask, well known to be highly selective to this chemistry, and a HSQ mask. Fluorine chemistries are commonly used to etch silicon but the formation of a passivation layer must be well-controlled in order to prevent chemical undercut. The mask was removed with a standard hydrofluoric  (HF) etch or a vapor HF etch, in the case of HSQ. The vertical etch flowed 130 sccm of C$_4$F$_8$ and 80 sccm of SF$_6$ at a pressure of 10 mTorr, RF power of 1000 W, and platen power of 12 W on a silicon carrier wafer. 

The same plasma parameters were used for the angled etch but a quartz wafer was used as the carrier. This reduces the loading due to Si etching and effectively frees more SF$_6$ molecules for etching. Because the angled etch of SF$_6$ radicals is directional while passivation layer formation is isotropic, a shadowing effect arises from the blocking of etching ions by the structure. In order to compensate for this to allow for a reasonable etching rate, more SF$_6$ must remain in the plasma, either by increasing the flow or switching carrier wafers. Using this recipe, silicon waveguides and high quality factor optical resonators have recently been demonstrated\cite{Latawiec2015b}.

\subsection{Quartz FCAE}

Quartz is an interesting material platform, widely used as a mechanical resonator in micro-electromechanical systems (MEMS) and other technologies\cite{Cady1922}. However, lacking a native thin-film technology, it is difficult to integrate into nano-scale systems. Motivated by its excellent material properties, we realized quartz nanobeam cantilever resonators via angled-etching\cite{Sohn2015} (Fig. \ref{EtchFig}(c)). A metal hard mask was patterned using a bi-layer poly(methyl methacrylate) (PMMA) liftoff procedure (cantilevers) or direct etching using an e-beam resist as a mask (optical resonators). The pattern was then transferred to the quartz using a CHF$_3$-based recipe at 10 mTorr pressure (STS MPX/LPX ICP-RIE). The ICP power was held at 600 W and the platen at 90 W. Argon, CF$_4$, CHF$_3$, and H$_2$ were all flowed at 6, 2, 50, and 15 sccm respectively. Due to the physical milling introduced by angled-etching and the small loadlock ($\sim$8 mm) on this etcher, the angled-etching was completed in a different machine which could accommodate a larger cage design (Nexx ECR RIE). This etcher relies on a different mechanism to generate a plasma (electron-cyclotron resonance), yet was still shown to be suitable for angled-etching. The microwave power was set to 600 W while the platen RF power was set to 90 W and the pressure was held at 10 mTorr. The ratio of gasses remained the same, with the overall flow rate reduced by a factor of two to accommodate machine constraints.

\section{Simulation}

\begin{figure}[!htbp]
\includegraphics{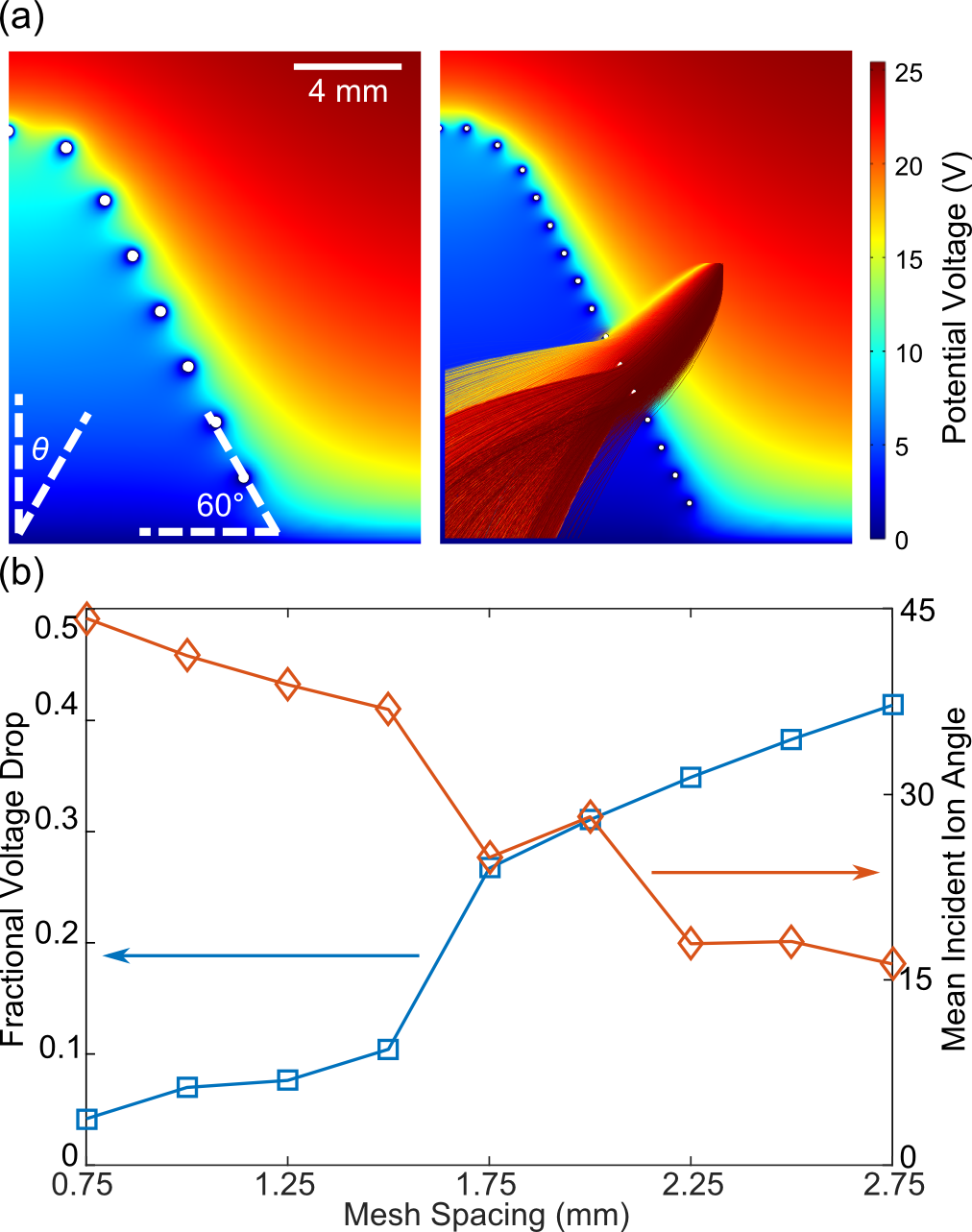}
 \caption{\label{VdropFig} (Color online) Simulated voltage drops inside the Faraday cage. (a) Simulated etching potentials for a coarse cage (left) and fine cage (right). The coarse cage has a larger potential gradient within, influencing the path the ion takes. The simulated ion trajectories from a single point outside the cage are overlayed on the right hand side, with coloring to provide visual contrast. All simulated cages have a cage angle of 60 degrees. (b) Voltage drop inside the cage relative to the etch potential (squares, left axis) and resultant mean incident ion etching angle, defined from the normal (diamonds, right axis). }
\end{figure}

The effect of the Faraday cage on the reactor potential was investigated with COMSOL Multiphysics using the plasma physics module\cite{Hsu2006}. This simulation technique has been shown to yield results of reaction parameters with reasonable agreement to experiment\cite{Hsu2006, Corr2008} and can be extended to couple to models describing feature profile evolution\cite{Hsu2008}. The modeling was preformed using a standard Gaseous Electronic Conference (GEC) reference cell with argon gas, a chemistry which qualitatively captures the impact of the presence of the Faraday cage. The simulation was performed under axial symmetry, with the cage wires forming rings around the sample. No biasing was coupled into the simulation, so the forward power remained at effectively 0 W. Figs. \ref{IntroFig}(a) and \ref{VdropFig}(a) show the potential distribution of the plasma with a cage placed inside. The cage wires are maintained at ground while the plasma potential develops in the simulation. The etcher frequency used in the simulation is the standard 13.56 MHz. The coil current is maintained at 100 A, and the power within is monitored until convergence (typically measured $\sim$2000 W). As expected, the plasma potential decreases as the plasma gets closer to the cage, creating electric fields normal to the cage face. Towards the bottom of the cage, the plasma sheath starts to resemble the sheath at the carrier wafer, flattening out. As the sheath moves up the cage, it follows its contours more closely, resulting in electric fields that are more normal to the cage face.

By changing the simulated cage geometry, we can investigate the effects of different cage types on the potential distribution using straightforward simulation parameters. In particular, we change the mesh parameters by changing the thickness of the wires or their spacing. Another parameter of interest is the sample height within the Faraday cage. This is changed by placing a metallic (grounded) block within the cage, and then recording the final ion velocities upon hitting the top surface, where the sample presumably lies. Throughout the simulations, the mesh face angle is maintained at 60 degrees. The impact of the mesh angle is not explicitly studied, as the etch angle has been seen to be proportional to the mesh face angle previously\cite{Burek2012b}.

To study the action of the ions under the cage potential, the motion of ions was simulated using a particle tracing module in COMSOL. After releasing the ions (argon) from the sheath with a Maxwellian velocity distribution (at a temperature of 400 K), their velocity and position is charted.  Once the ion intersects with a defined sample region, its velocity vector and position is recorded to generate an angle/energy ion distribution map. This simulation procedure is repeated with the different cage geometries, showing stark differences in the incident angle distribution, as well as ion energies.

\begin{figure}[!htbp]
\includegraphics{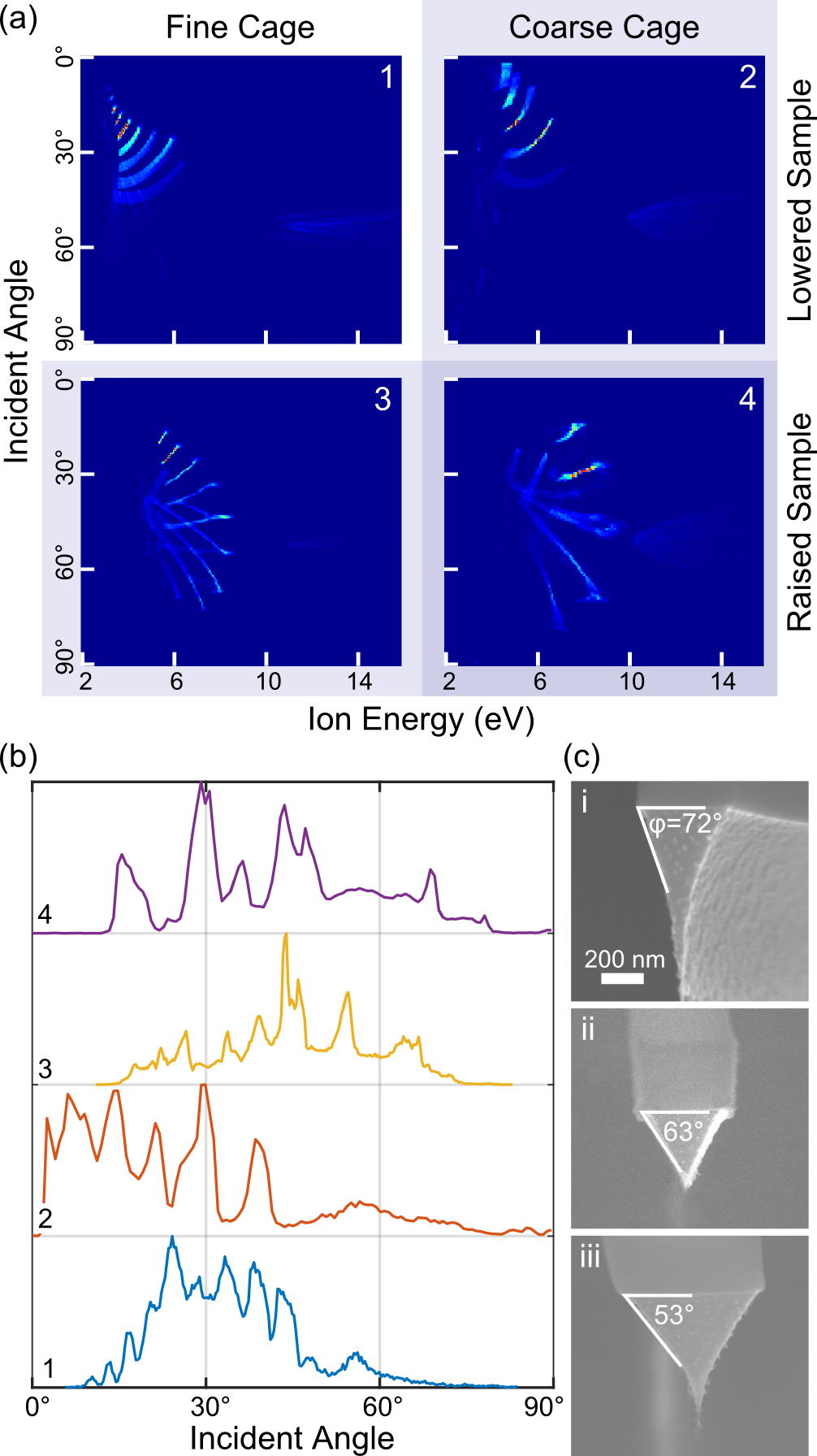}
 \caption{\label{HistFig} (Color online) Detailed results from changing cage parameters. (a) Incident angle/ion energy histograms for simulated cage designs with fine (left) and coarse (right) mesh spacings and raised (bottom) and lowered (top) samples. The ion data are taken upon collision with the sample area. The angle-energy histograms show clear "banding" of the ions due to the effect of the Faraday cage wires. Notably, both the raised sample position and coarse mesh spacing increase the variability in angle/momentum space.  Generally, higher ion energy (equivalently, momenta) can negatively effect mask selectivity as the etch becomes more physical. (b) Averaged ion incident angle for the cage designs in part (a), corresponding to their number. (c) Head-on SEM images of cantilevers etched in (i) a raised, coarse cage, (ii), a lowered, fine cage, and (iii) a raised, fine cage. The angle $\varphi$ is defined as the observed etch angle. The SEMs show that $\varphi$ varies in accordance with simulation. The visible sidewall roughness seems to be a characteristic of the particular etch and is not greatly affected by cage design.}
\end{figure}

\begin{figure}[!htbp]
\includegraphics{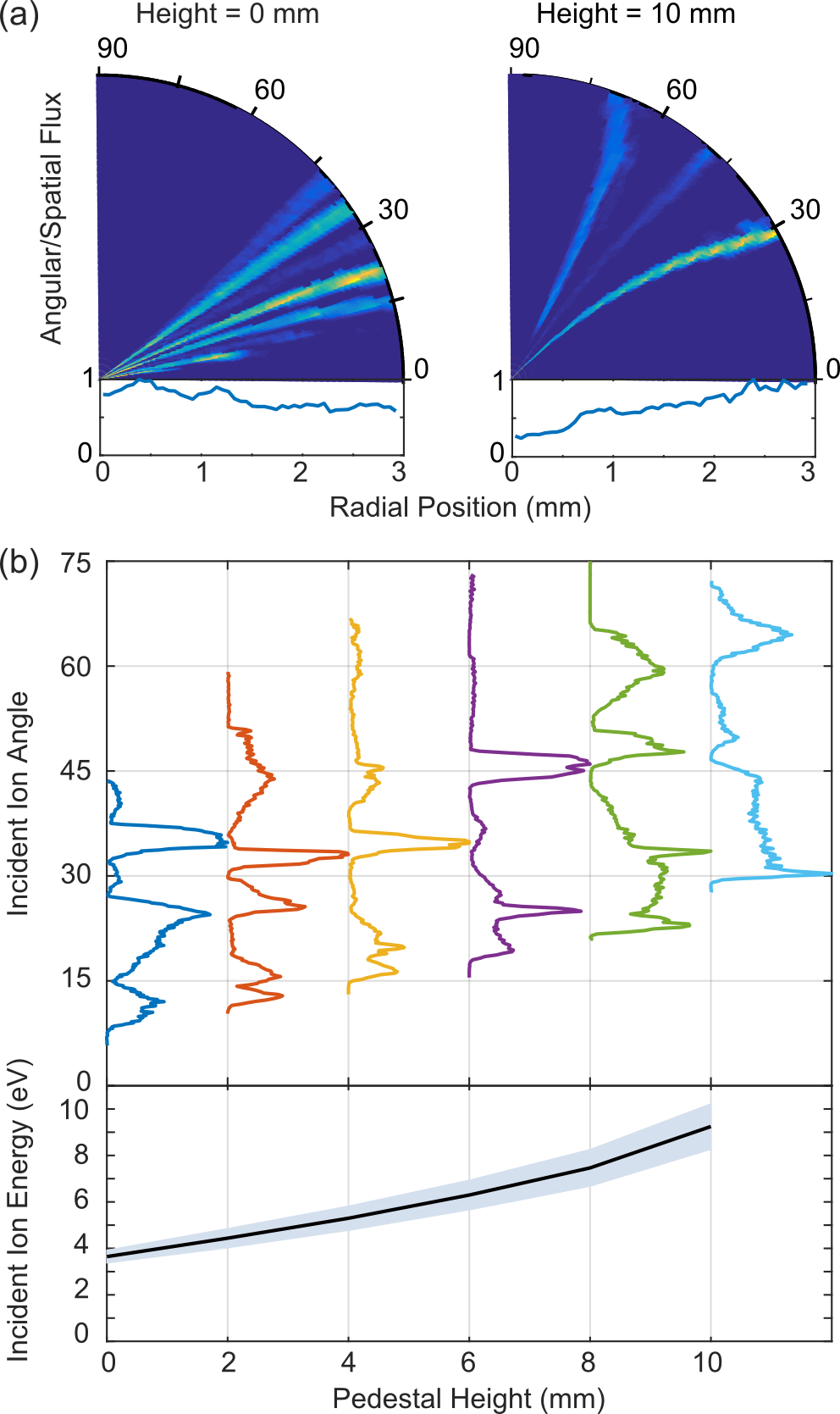}
 \caption{\label{SpatialFig} (Color online) Impact on sample height and incident angle across a sample. (a) Incident ion angle as a function of radial position on the sample for samples on the carrier (left) and placed on a 10 mm pedestal (right). The bottom shows the relative flux of etching ions as a function of position. The sample placed on the pedestal shows a larger variation in ion flux and incident angle. (b) Incident ion angles for a cage as the sample is placed on higher pedestals (top). Mean ion energy and standard deviation for ions as the sample height is raised (bottom) }
\end{figure}

\section{Results and Discussion}

Synthesizing the simulation results and incorporating experimental evidence, we can make definite statements on the plasma physics and ion dynamics, whose qualitative ideas transfer to the etching of many other material types in FCAE. To begin with, the simulations (Fig. \ref{VdropFig}) show that coarser meshes allow the leakage of potential past the mesh, resulting in a secondary ion sheath inside the cage. This implies the existence of a secondary electric field pointed normal to the carrier wafer, after the initial electric field which accelerates the ions and gives them their angle. This secondary field is detrimental to the function of the cage, adding a vertical component to the incident ion angle. In Fig. \ref{VdropFig}(a), the simulated cage potentials show the elevated internal potential in the coarsely-spaced mesh. Fig. \ref{VdropFig}(b) makes the link between the internal secondary sheath and ion incident angle explicit. As the mesh spacing increases, the fractional voltage drop (defined as the  maximum voltage within the cage divided by the maximum voltage outside the cage) increases to as high as 40\%. Likewise, as this internal voltage drop increases, the average incident ion angle decreases, as the ions are deflected to a more vertical incidence (here defined as zero degrees).

By looking at the ion distribution in both incident angle and energy, as in Fig. \ref{HistFig}(a), we can fine-tune our observations. An ideal Faraday cage design would show a narrow distribution of ions in both angle and energy space, implying that all incident ions are similar in both direction and momentum, providing a well-defined and controlled etch. In reality, we see that finer meshes are seen to create a larger incident etching angle on the surface, closer to the value prescribed by the cage geometry, although they also create tightly-defined “bands” of angles. These bands are a result of the local deflection of the trajectory of the ions due to the mesh wire itself\cite{Cho}. Finer meshes have more wires per unit area, resulting in more of these bands. However, the larger wire sizes of coarser meshes tend to bend the trajectory for a longer time, increasing the glancing angle. To simplify the interpretation of these results, we project the ion distribution onto the angular axis, averaging over the different ion energies, as in Fig. \ref{HistFig}(b). The corresponding distributions carry the same label as in part (a). These simplified charts show the decrease in standard deviation of incident angle afforded by finer meshes (labels 1 and 3). They also simultaneously show the increase in mean ion etch angle as the sample is raised within the Faraday cage, an effect which is further explored in Fig. \ref{SpatialFig}. In a real chamber, a DC bias is expected to increase the electric potential gradient, resulting in a mean incident ion angle closer to that prescribed by the Faraday cage. The stronger electric fields should also reduce the perturbation on the ion trajectories caused by the cage wires, resulting in narrower ion distributions.

To support the results from these simulations, silicon cantilever samples were fabricated with the recipe outlined beforehand and imaged at head-on incidence to record the etch angle, defined as $\varphi$ in Fig. \ref{HistFig}(c). A coarse mesh was observed to increase the verticality of the etch (larger $\varphi$, Fig. \ref{HistFig}(c,i)) even though the sample was raised. Positioning a sample lower while using a fine mesh made the etch more vertical (Fig. \ref{HistFig}(c,ii), while raising the sample within a fine mesh gave the most acute etch angle, Fig. \ref{HistFig}(c,iii). This corroborates exactly the results from simulation, where the incident angle changes starkly based on sample positioning and mesh type (Fig. \ref{HistFig}(a,b)).

The etch rate and etch angle at different points inside the conical Faraday cage was experimentally shown to vary dramatically due to the asymmetry of the cage itself and off-center placement of the sample within. In order to mitigate the effects of this so-called etch gradient, the sample was etched at intervals of 30 seconds, with rotations of the cage in-between\cite{Burek2015}. Although this process significantly symmetrizes the etch angle across the device, the etch rate was still observed to be variable, with the highest etch rate in the middle of the device. This is supported by the simulations of the ion trajectories in Fig \ref{SpatialFig}(a). Toward the center of the sample, there is a large flux of ions coming not only from the side of the cage (at an angle), but also from the top of the cage, where it curves to become horizontal (directing ions downward). Capping the top of the cage with a shield can prevent these stray ions from entering the cage. Experimentally, it was also observed that cages which resulted in more acute etch angles $\varphi$ had larger etch gradients. This can also be observed in the simulation data, as the angle/ion energy histogram displays a large spread of incident ion angle for samples positioned higher and with tighter meshes. Looking across the sample, Fig. \ref{SpatialFig}(a) shows just how sample position affects ion properties. For low-positioned samples, the angle distribution of ions is constant across the sample. There is some modulation of flux, as seen in the bottom chart, possibly due to the effects mentioned beforehand. Highly-positioned samples show a larger variation in flux and a large change in incident ion angle, especially towards the edge of the sample. In general, from Fig. \ref{SpatialFig}(b), it can be seen that higher-positioned samples tend to not only have higher-energy incident ions, but a larger spread of incident angles and energies as well, possibly leading to less-controlled etching conditions.

It is a well-known result from ion-beam etching that physical mask milling changes with changes in etch angle. Within FCAE etching, the selectivity of the etch decreases as more acute etch angles ($\varphi$ approaching 0) are attempted\cite{Boyd1980}. This effect is seen qualitatively in diamond devices using both HSQ and alumina as a mask. Furthermore, because of the milling action that occurs when etching ions are incident on the mask, acute etch angles are also associated with an increase in micromasking, roughening the surface of the etched device significantly. Likewise, increasing the mask coverage of the sample leads to an increase in micromasking as more mask particles are resputtered into the chamber. This can be mitigated by designing patterns with less mask area.

With this in mind as well as the results from Fig. \ref{SpatialFig}(a), angled-etching may also be used to non-invasively engineer continuous etch profiles on a sample surface. This can be accomplished by appropriately shaping the design of the Faraday cage. In our simulations, we have shown that even a cage without bends (and therefore a nominally constant radius of curvature equal to zero) can create a spatially-varying angular and flux distribution in ions, thereby modifying the local etch rate. Adding curved components in the Faraday cage can create a lensing effect, in analogy with electron and ion optics. This can be used to three-dimensionally pattern a surface without any lithographic steps\cite{Kovi2015}.

\section{Summary and Conclusion}

Faraday cage angled-etching has been demonstrated to be a robust platform with which to etch a wide range of materials\cite{Burek2012b, Yamaguchi2015, Latawiec2015b}, marrying the high material quality of bulk single-crystal substrates with the versatility of a three dimensional etching technique able to etch complex nanostructures. After reviewing the design of Faraday cages suitable for etching in standard ICP-RIE tools and looking at results in three widely different material platforms (diamond, silicon, and quartz), we studied the influence of cage parameters on the etch angle, uniformity, and selectivity. Simulations of FCAE within a standard reactor setup helped illuminate the dynamics involved in angled-etching, driving physical intuition. These results should help guide future effort in nanofabrication using FCAE, as well as inform new paths for extensions to this technology. This study could be expanded by looking into the effects of different plasma chemistry and forward bias on the plasma potential, as well as by incorporating a full 3D model of the etching chamber and Faraday cage. 

%
%

%

\begin{acknowledgments}
This work was performed in part at the Center for Nanoscale Systems (CNS), a member of the National Nanotechnology Infrastructure Network (NNIN), which is supported by the National Science Foundation under NSF award no. ECS-0335765. CNS is part of Harvard University. This work was supported by the DARPA SCOUT program through grant number W31P4Q-15-1-0013 from AMRDEC. P. L. is supported by the National Science Foundation Graduate Research Fellowship under Grant No. DGE1144152. M. J. B. was supported in part by the Natural Science and Engineering Council (NSERC) of Canada and the Harvard Quantum Optics Center (HQOC).
\end{acknowledgments}

\bibliography{EIBPN}

\end{document}